\documentclass[]{aastex63}

\accepted{ApJ Letters, \today}
\shorttitle{Turbulence in the Sub-Alfv\'enic Solar Wind}
\shortauthors{Zank et al.}

\begin{document}

\title{Turbulence in the Sub-Alfv\'enic Solar Wind}

\correspondingauthor{G.P.\ Zank}
\email{garyp.zank@gmail.com}

\author[0000-0002-4642-6192]{G.P.\ Zank}
\affiliation{Center for Space Plasma and Aeronomic Research (CSPAR) and Department of Space Science, University of Alabama in Huntsville, Huntsville, AL 35805, USA}

\author{L.-L. Zhao}
\affiliation{Center for Space Plasma and Aeronomic Research (CSPAR) and Department of Space Science, University of Alabama in Huntsville, Huntsville, AL 35805, USA}

\author{L. Adhikri}
\affiliation{Center for Space Plasma and Aeronomic Research (CSPAR) and Department of Space Science, University of Alabama in Huntsville, Huntsville, AL 35805, USA}

\author{D. Telloni}
\affiliation{INAF—Astrophysical Observatory of Torino, Via Osservatorio 20, I-10025 Pino Torinese, Italy}

\author{J. C. Kasper}
\affiliation{BWX Technologies, Inc., Washington DC 20002, USA and 
	Department of Climate and Space Sciences and Engineering, University of Michigan, Ann Arbor, MI 48109, USA}
	
\author{M. Stevens}	
\affiliation{Smithsonian Astrophysical Observatory, Cambridge, MA 02138 USA}

\author{A. Rahmati}	
\affiliation{Space Sciences Laboratory, University of California, Berkeley, CA 94720-7300, USA}

\author{S. D. Bale}
\affiliation{Physics Department, University of California, Berkeley, CA 94720-7300, USA}



\begin{abstract}
\textit{Parker Solar Probe (PSP)} entered a region of sub-Alfv\'enic solar wind during encounter 8 and we present the first detailed analysis of low-frequency turbulence properties in this novel region. The magnetic field and flow velocity vectors were highly aligned during this interval. By constructing spectrograms of the normalized magnetic helicity, cross helicity, and residual energy, we find that \textit{PSP} observed primarily Alfv\'enic fluctuations, a consequence of the highly field aligned flow that renders quasi-2D fluctuations unobservable to \textit{PSP}. We extend Taylor's hypothesis to sub- and super-Alfv\'enic flows. Spectra for the fluctuating forward and backward Els\"asser variables (${\bf z}^{\pm}$ respectively) are presented, showing that ${\bf z}^+$ modes dominate ${\bf z}^-$ by an order of magnitude or more and the ${\bf z}^+$ spectrum is a power law in frequency (parallel wave number) $f^{-3/2}$ ($k_{\parallel}^{-3/2}$) compared to the convex ${\bf z}^-$ spectrum with $f^{-3/2}$ ($k_{\parallel}^{-3/2}$) at low frequencies, flattening around a transition frequency (at which the nonlinear and Alfv\'en time scales are balanced) to $f^{-1.25}$ at higher frequencies. The observed spectra are well fitted using a spectral theory for nearly incompressible magnetohydrodynamics assuming a wave number anisotropy $k_{\perp} \sim k_{\parallel}^{3/4}$, that the ${\bf z}^+$ fluctuations experience primarily nonlinear interactions, and that the minority ${\bf z}^-$ fluctuations experience both nonlinear and Alfv\'enic interactions with ${\bf z}^+$ fluctuations. The density spectrum is a power law that resembles neither the $\mathbf{z^{\pm}}$ spectra nor the compressible magnetic field spectrum, suggesting that these are advected entropic rather than magnetosonic modes and not due to the parametric decay instability. Spectra in the neighboring modestly super-Alfv\'enic intervals are similar.
\end{abstract}

\keywords{solar corona, turbulence}


\section{Introduction}

For about 5 hours between 09:30--14:40 UT on 2021-04-28 at around 0.1 au, the NASA {\textit{Parker Solar Probe (PSP)} entered a sub-Alfv\'enic region of the solar wind \citep{Kasper_etal_2021}. Two further shorter sub-Alfv\'enic intervals were subsequently sampled during encounter 8. \cite{Kasper_etal_2021} ascribe the first sub-Alfv\'enic region to a steady flow in a region of rapidly expanding magnetic field above a pseudostreamer. The discovery of this hitherto in situ unobserved region of the solar wind represents a major accomplishment of the \textit{PSP} mission, particularly for the insight it will provide in our understanding of how the solar corona is heated and the solar wind accelerated. The dissipation of low frequency turbulence is regarded as a promising mechanism for heating the solar corona. The current explicitly turbulence models come in essentially two flavors, one dominated by outwardly propagating Alfv\'en waves, a sufficient number of which are reflected by the large-scale coronal plasma gradient to produce a counter-propagating population of Alfv\'en waves that interact nonlinearly to produce zero frequency modes that cascade energy nonlinearly to the dissipation scale to heat the corona \citep{matthaeus_etal_1999_coronalheating, Verdini_etal_2009, Cranmer_vanBallegooijen_2012, Shoda_etal_2018, Chandran_Perez_2019}. The second approach recognizes that magnetohydrodynamics (MHD) in the plasma beta $\beta_p \equiv P/(B^2/2 /\mu_0) \ll 1$ or $O(1)$ regimes ($P$ the plasma pressure, $B = |{\bf B}|$, ${\bf B}$ the magnetic field, and $\mu_0$ the magnetic permeability) is quasi-2D at leading order \citep{Zank_Matthaeus_1992_solarwind, zank_matthaeus_1993_incompress}, with the result that turbulence in these regimes is dominated by quasi-2D turbulence with a minority slab turbulence component \citep{Zank_etal_2017a}. Nearly incompressible MHD (NI MHD) is the foundation of the well-known 2D+slab superposition model for turbulence in the solar wind \citep{matthaeus_etal_1990_quasi2dfluct, Bieber_etal_1994, Bieber_etal_1996JGR}. The NI MHD description forms the basis of the coronal turbulence heating model advocated by \cite{Zank_etal_2018} for which a dominant population of turbulent MHD structures (flux ropes/magnetic islands, vortices, plasmoids) is generated in the magnetic carpet of the photosphere and advected through and dissipated in the low corona. Accompanying the majority quasi-2D turbulence is a minority population of Alfv\'enic or slab turbulence, most likely predominantly outward propagating. A comparative analysis of the two turbulence models using \textit{PSP} observations is presented in \cite{Zank_etal_2021b}. Here we examine the properties of low-frequency MHD turbulence in the first sub-Alfv\'enic interval observed by \textit{PSP} and show that these observations admit a natural interpretation in terms of the NI MHD spectral theory \citep{Zank_etal_2020a}. 

\begin{figure}[htbp]
\begin{center}
\includegraphics[width= 1.0\textwidth]{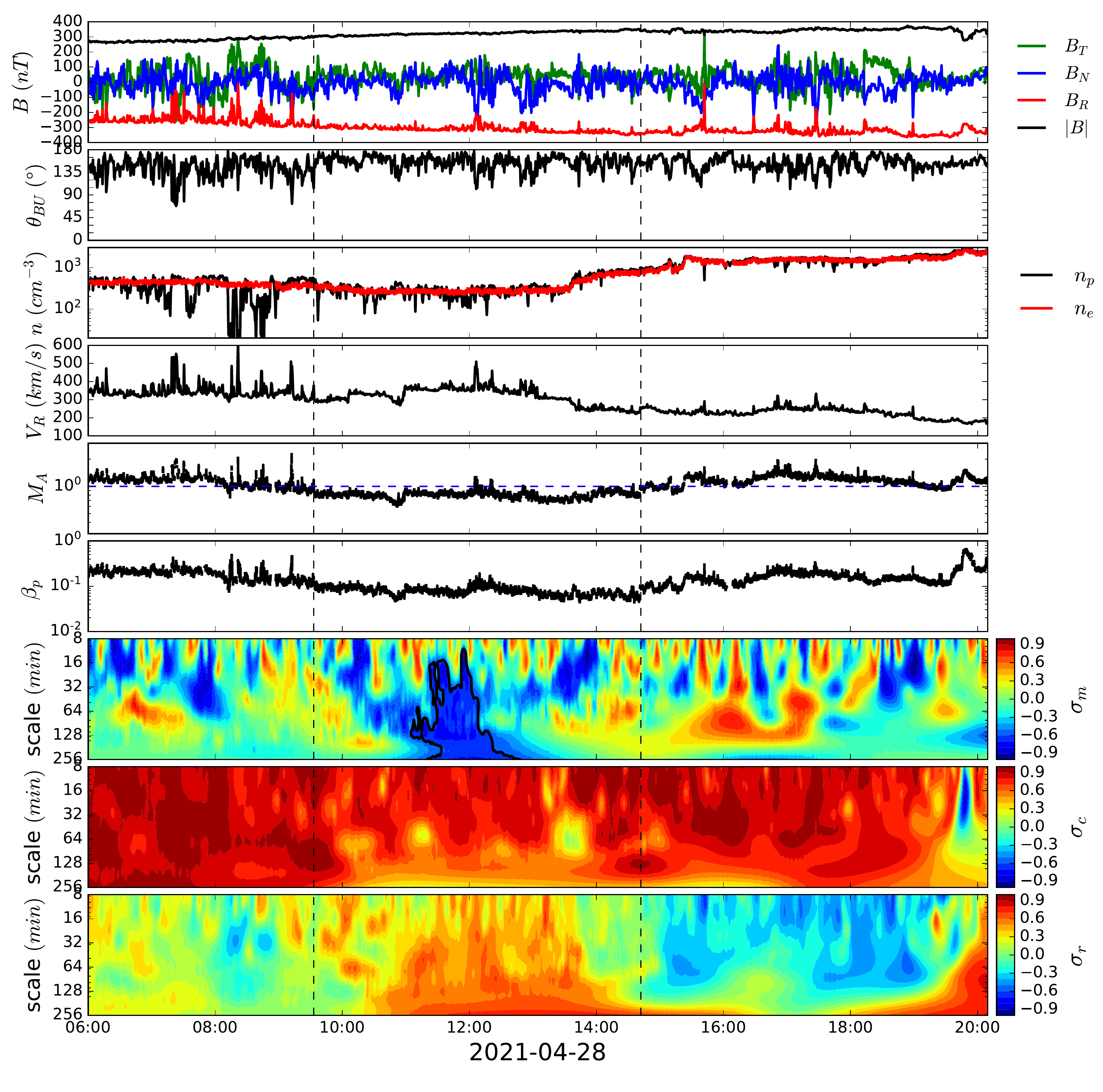}
\end{center}
\caption{An overview of the first sub-Alfv\'enic interval, located between the dashed vertical lines, and the adjacent super-Alfv\'enic intervals observed by \textit{PSP} during encounter 8 showing from top to bottom the radial $B_R$, transverse $B_T$, and normal $B_N$ magnetic fields, and the intensity $|{\bf B}|$, the angle $\theta_{BU}$ between the relative flow velocity ${\bf U}_0$ (i.e., the relative velocity between the solar wind flow and the spacecraft velocity vectors) and magnetic field vectors, the proton $n_p$ (black) and electron number density $n_e$ (red),  the radial component of the plasma speed $V_R$ measured in the inertial RTN frame, Alfv\'enic Mach number of the radial flow $M_A = V_R /V_A$, and the plasma beta $\beta_p$. The bottom three colored panels, in descending order, show frequency spectrograms of the normalized magnetic helicity $\sigma_m$, normalized cross helicity $\sigma_c$, and normalized residual energy $\sigma_r$.  }
\label{fig:1}
\end{figure}

Figure \ref{fig:1} is an overview of the first and longest of three sub-Alfv\'enic intervals identified by \cite{Kasper_etal_2021}. The data we used include magnetic field measurements from PSP/FIELDS \citep{Bale_etal_2016}, ion moments data from PSP/SWEAP instrument, and electron density derived from quasi-thermal noise (QTN) spectroscopy \citep{Kasper_etal_2016, Kasper_etal_2021}. The radial magnetic field is extremely steady with relatively small amplitude fluctuations. The radial velocity is also rather steady and the alignment between the flow and magnetic field vectors is very high. Below, we introduce the quantity $\Psi \equiv \langle \theta_{B_0 U_0} \rangle$ where ${\bf B}_0$ and ${\bf U}_0$ are the magnetic and plasma velocity mean fields during the intervals of interest. The velocity ${\bf U}_0$ is the relative velocity that includes the spacecraft speed, i.e., the spacecraft frame velocity. As we discuss in Section \ref{sec:frequency}, this is important for the generalized form of Taylor's hypothesis that we use. The flow appears to be mostly sub-Alfv\'enic, although relatively marginally, across much of the interval, since the Alfv\'enic Mach number $M_{A}=V_{R}/V_{A} \lesssim 1$. That the flow is so highly aligned renders quasi-2D fluctuations essentially invisible to \textit{PSP}, as we show explicitly below, and only fluctuations propagating along or anti-parallel to the inwardly directed (towards the Sun) magnetic field are observable. The plasma beta $\beta_p$ is well below 1, being close to $10^{-1}$ for most of the interval. In the NI MHD context, this would imply that the predicted majority quasi-2D component was not observed because of the close alignment of the flow with the magnetic field; instead the observed fluctuations correspond to a minority slab component. 

We have developed a method \citep{Zhao2019, Zhao_etal_2020} to automatically identify magnetic flux ropes and Alfv\'enic fluctuations based on the observed rotation of the magnetic field \citep{Burlaga1981, Moldwin1995} and the normalized reduced magnetic helicity \citep{Matthaeus_etal_1982}, which is usually high  in regions of magnetic flux ropes \citep{Telloni2012, Telloni2013}. To distinguish between Alfv\'enic structures and flux ropes, we evaluate the normalized cross helicity $\sigma_c$ ($\equiv (\langle {z^+}^2 \rangle - \langle {z^-}^2 \rangle )/( \langle {z^+}^2 \rangle + \langle {z^-}^2 \rangle)$, where ${\bf z}^{\pm} = {\bf u} \pm {\bf b}/\sqrt{\mu_0 \rho}$ are the Els\"asser variables for the fluctuating velocity ${\bf u}$ and magnetic ${\bf b}$ fields and $\rho$ the mean plasma density,  and the normalized residual energy $\sigma_r$ ($\equiv \langle {\bf z}^+ \cdot {\bf z}^- \rangle /( \langle {z^+}^2 \rangle + \langle {z^-}^2 \rangle) =  (\langle u^2 \rangle - \langle b^2 \rangle / (\mu_0 \rho) )/ (\langle u^2 \rangle + \langle b^2 \rangle /(\mu_0 \rho))$). In most cases, the cross helicity of a magnetic flux rope is low (because of their closed-loop field structure, implying both sunward and anti-sunward Alfv\'enic fluctuations inside) and the residual energy is negative (indicating the dominance of magnetic fluctuation energy), whereas Alfv\'enic structures typically exhibit high cross helicity and a small residual energy values. Frequency spectrograms of the normalized magnetic helicity $\sigma_m$, normalized cross helicity $\sigma_c$, and normalized residual energy $\sigma_r$ are illustrated in the bottom three panels of Figure \ref{fig:1}. Several points are immediately apparent. There are numerous small and mid-scale magnetic positive and negative rotations, including a particularly large structure bounded by black contour lines with $\langle \sigma_m \rangle  \simeq -0.7$. Within this high magnetic helicity structure, the averaged $\sigma_c$ is around $0.55$ and the averaged $\sigma_r$ is about $0.5$. The scale size of this structure is around 80 minutes. The plasma outside this large structure has a cross helicity, close to 1,  indicating almost exclusively outward propagating anti-parallel to the mean magnetic field, Els\"asser fluctuations ${\bf z}^{+}$. The residual energy spectrogram shows large regions with $\sigma_r \simeq 0$, i.e., Alfv\'enic fluctuations, although the border region near 16:00 hours, 2021-04-28, appears to be comprised of largely magnetic structures with with negative $\sigma_r$. The left-handed helical structure ($\sigma_m < 0$) at about 12:00 has a positive $\sigma_r$ ($\sim 0.5$), indicating the relative dominance of kinetic fluctuating energy. It suggests that \textit{PSP} may have observed a vortical structure with a relatively weak wound-up magnetic field. 

The combination of $\sigma_c \sim 0.9$ and $\sigma_r \sim 0.0$ for much of the sub-Alfv\'enic interval suggests that, not surprisingly, \textit{PSP} is observing primarily Alfv\'enic fluctuations. This is typically described as ``Alfv\'enic turbulence'' despite the inability of \textit{PSP} to discern non-Alfv\'enic structures easily in a highly magnetic field-aligned flow. The very high value of $\sigma_c$ raises considerable problems for slab turbulence, which relies on counter-propagating Alfv\'en waves to generate the non-linear interactions that allow for the cascading of energy to ever smaller scales  \citep{Dobrowolny_etal_1980a, Dobrowolny_etal_1980b, Shebalin_etal_1983}. Such highly field-aligned flows with high $\sigma_c$ and populated by uni-directionally propagating Alfv\'en waves have been observed near 1 au \citep{Telloni_etal_2019, wang2015} by the \textit{Wind} spacecraft and closer to the Sun by \textit{PSP} \citep{Zhao_etal_2020a, Zhao_etal_2021}. The 1D reduced spectra in both cases exhibited a $k_{\parallel}^{-5/3}$ form in the inertial range, ($k_{\parallel}$ the wave number parallel to the mean magnetic field), raising questions about the validity of  critical balance theory  \citep{Goldreich_Sridhar_1995, Telloni_etal_2019}. The spectral theory of NI MHD in the $\beta_p \ll 1$ and $\sim 1$ regimes \citep{Zank_etal_2020a} shows that the interaction of a dominant quasi-2D component with uni-directional Alfv\'en waves  can yield a $k_{\parallel}^{-5/3}$ spectrum. The 
sub-Alfv\'enic interval of Figure \ref{fig:1} exhibits a number of features quite similar to those observed in the highly field-aligned flows discussed by \cite{Telloni_etal_2019} and \cite{Zhao_etal_2020a, Zhao_etal_2021}. 

In this Letter, we analyze the turbulent properties of the sub-Alfv\'enic interval shown in Figure \ref{fig:1} in more detail than done in \cite{Kasper_etal_2021} and interpret the observations based on the NI MHD spectral theory appropriate to the anisotropic superposition of quasi-2D+slab turbulence \citep{Zank_etal_2020a}. This does however require the correct application of Taylor's hypothesis for the sub-Alfv\'enic and modestly super-Alfv\'enic flows discussed here. Such an extension is  straightforwardly developed in the context of the NI MHD superposition model, but it does require that the ${\bf z}^{\pm}$ Els\"asser modes be treated separately for the minority forward and backward propagating slab modes. This is done in the following section, after which we apply the results to the ${\bf z}^{\pm}$ spectra observed by \textit{PSP}.

\section{Relating Wave Number and Frequency Spectra}   \label{sec:frequency}

Taylor's hypothesis assumes that a simple Galilean transformation $\omega = {\bf U}_0 \cdot {\bf k}$,  can relate the wave number ${\bf k}$ in the inertial frame to the observed frequency ($\omega$ or $2\pi f$). Here, we assume implicitly that ${\bf U}_0$ is the relative velocity between the background solar wind flow velocity and the spacecraft velocity. This is reasonable in the fully developed supersonic solar wind where one can assume that characteristic wave speeds are $\ll {\bf U}_0$, particularly the Alfv\'en velocity that satisfies $|{\bf V}_A \cdot {\bf k}| \ll |{\bf U}_0 \cdot {\bf k}|$, but Taylor's hypothesis is unlikely to be appropriate to sub-Alfv\'enic or modestly super-Alfv\'enic regions of the solar wind. Suppose that ${\bf x} = \left( {V}_p + {\bf U}_0 \right) t = {\bf V}t$, ${\bf x}^{\prime} = {\bf x} + \mbox{\boldmath$\ell$}$, $t^{\prime} = t + \tau$, where {\boldmath$\ell$} and $\tau$ are spatial and temporal separations from ${\bf x}$ and $t$, and ${\bf V}_p$ denotes a phase velocity.  On making the usual assumptions of homogeneity and stationarity, it is straightforward to show that the power spectral density (PSD) is \citep{Bieber_etal_1996JGR, Sauer_Bieber_1999, Zank_etal_2020a}
\begin{equation}
P_{ij} (f) = \int e^{i 2\pi f \tau} \int P_{ij} ({\bf k}) e^{-i {\bf k} \cdot {\bf V} \tau  }d{\bf k} d\tau . \label{eq:1}
\end{equation}
For 2D modes that experience only advection, $\omega = 0$ whereas for slab/Alfv\'enic turbulence $\omega = \pm V_A k_{\parallel} = \pm V_A k_z$. For 2D turbulence, since $V_p = 0$, we have simply 
\begin{equation}
P_{ij}^{\infty} (f) = \int e^{i 2\pi f \tau} \int P_{ij}^{\infty} ({\bf k}) e^{-i {\bf k} \cdot {\bf U}_0 \tau } d{\bf k} d\tau . \label{eq:2}
\end{equation}
For slab turbulence, we need to decompose the fluctuations into forward ($+$) and backward ($-$) propagating modes (some care needs to be exercised in practice relative to the mean magnetic field orientation). On extending \cite{Zank_etal_2020a}, the slab PSD is given by 
\begin{equation}
P_{ij}^{* \pm} (f) = \int e^{i 2\pi f \tau} \int P_{ij}^{* \pm} ({\bf k}) e^{-i (\mp V_A k_z + {\bf U}_0 \cdot {\bf k}) \tau} d{\bf k} d\tau , \label{eq:3}
\end{equation}
with the $\mp V_A$ appearing in the exponential because formally the ${\bf z}^{* +}$ transport  equation contains the advection term $({\bf U} - {\bf V}_A) \cdot \nabla$ and the ${\bf z}^{* -}$ transport  equation has $({\bf U} + {\bf V}_A) \cdot \nabla$ \citep{Zank_etal_2017a}. Other approaches to Taylor's hypothesis that are not based on an assumed 2D+slab decomposition of the turbulence have been considered \citep[e.g.,][]{Klein_etal_2015, Bourouaine_Perez_2020}. We comment that although  the generalized form of Taylor's hypothesis used here (based on a 2D + slab decomposition) accounts for the  ${\bf z}^+$ component (propagating at phase speed of ${\bf U}-{\bf V}_A$) and the  ${\bf z}^-$ component  associated with phase speed (${\bf U} + {\bf V}_A$), we do not account for the ``anomalous'' ${\bf z}^-$ component arising from reflection and ``mixing'' from large-scale flow gradients and ${\bf z}^+$.    

Following \cite{Bieber_etal_1996JGR}, we approximate $P_{ij}^{\infty,*\pm}({\bf k})$ by  general forms of the 2D (``$\infty$'') and forward and backward slab (``$*\pm$'') spectral tensors (equations (4) and (5) in \cite{Zank_etal_2020a}) \citep{Matthaeus_Smith_1981}. The spectral theory developed in \cite{Zank_etal_2020a} is for the Els\"asser variables, and we therefore focus on the Els\"asser representation for $P_{ij}^{\infty,*\pm}({\bf k})$ here. It is reasonable to assume isotropy for the 2D fluctuations. On performing the suitable integrals, the 2D results of \cite{Zank_etal_2020a} are unchanged (assuming isotropy), whereas for super-Alfv\'enic flows satisfying $U_0 \cos \Psi > |V_A|$ (where the local mean magnetic field defines the $\hat{\bf z}$-axis, the $\hat{\bf x}$-axis by the planes of the mean magnetic field ${\bf B}_0$ and ${\bf U}_0$, $\hat{\bf y}$ is orthogonal to the $\hat{\bf x}$-$\hat{\bf z}$-plane, and $\Psi$ the angle between ${\bf B}_0$ and ${\bf U}_0$), 
\begin{equation}
P_{xx}^{*\pm} (f) = P_{\parallel}^{*\pm} (f) = \frac{C^{*\pm} }{2} \frac{2\pi}{U_0 \cos \Psi \mp V_A} G^{*\pm} (k_z) = P_{yy}^{*\pm} (f) = P_{\perp}^{*\pm} (f), \quad k_z = \frac{2\pi f}{U_0 \cos \Psi \mp V_A}, \label{eq:4}
\end{equation}
where $C^{*\pm}$ are the amplitudes of the forward and backward propagating slab fluctuations. Here, $G^{*\pm} (k_z)$ is the spectral expression for slab turbulence in the NI MHD $\beta_p \ll 1$, $\sim 1$ regime and was derived in \cite{Zank_etal_2020a} and is discussed below. For sub-Alfv\'enic flows, we find
\begin{equation}
P_{xx}^{*\pm} (f) = P_{\parallel}^{*\pm} (f) = \frac{C^{*\pm} }{2} \frac{2\pi}{V_A \mp U_0 \cos \Psi } G^{*\pm} (k_z) = P_{yy}^{*\pm} (f) = P_{\perp}^{*\pm} (f), \quad k_z = \frac{2\pi f}{V_A \mp U_0 \cos \Psi}. \label{eq:5}
\end{equation}
Of course, $P_{total}^{*\pm} (f) = P_{\parallel}^{*\pm} (f) + P_{\perp}^{*\pm} (f)$.

The composite spectra for super-Alfv\'enic and sub-Alfv\'enic flows are therefore given by (\ref{eq:6}) and (\ref{eq:7}), and (\ref{eq:8}) and (\ref{eq:9}) respectively below, 
\begin{eqnarray}
P_{\parallel}^{total \pm} (f) = P_{\parallel}^{\infty} (f) + P_{\parallel}^{* \pm} (f) = \frac{C^{\infty} }{q^{\infty} + 1} \left( \frac{U_0 \sin \Psi }{2\pi} \right)^{q^{\infty} -1} f^{-q^{\infty}} + \frac{C^{*\pm} }{2} \frac{2\pi}{U_0 \cos \Psi \mp V_A} G^{*\pm} (k_z) ; \label{eq:6} \\
P_{\perp}^{total \pm} (f) = P_{\perp}^{\infty} (f) + P_{\perp}^{* \pm} (f) = \frac{q^{\infty} }{q^{\infty} + 1} C^{\infty} \left( \frac{U_0 \sin \Psi }{2\pi} \right)^{q^{\infty} -1} f^{-q^{\infty}} + \frac{C^{*\pm} }{2} \frac{2\pi}{U_0 \cos \Psi \mp V_A} G^{*\pm} (k_z), \label{eq:7}
\end{eqnarray}
and $k_z = 2\pi f /(U_0 \cos \Psi \mp V_A)$, $C^{\infty}$ the amplitude of the 2D turbulence, and $q^{\infty}$ the spectral index of the 2D component; and 
\begin{eqnarray}
P_{\parallel}^{total \pm} (f) =  \frac{C^{\infty} }{q^{\infty} + 1} \left( \frac{U_0 \sin \Psi }{2\pi} \right)^{q^{\infty} -1} f^{-q^{\infty}} + \frac{C^{*\pm} }{2} \frac{2\pi}{V_A \mp U_0 \cos \Psi } G^{*\pm} (k_z) ; \label{eq:8} \\
P_{\perp}^{total \pm} (f) = \frac{q^{\infty} }{q^{\infty} + 1} C^{\infty} \left( \frac{U_0 \sin \Psi }{2\pi} \right)^{q^{\infty} -1} f^{-q^{\infty}} + \frac{C^{*\pm} }{2} \frac{2\pi}{V_A \mp U_0 \cos \Psi } G^{*\pm} (k_z), \label{eq:9}
\end{eqnarray}
and $k_z = 2\pi f /(V_A \mp U_0 \cos \Psi)$. The spectral theory for $\beta_p \ll 1$, $\sim 1$ NI MHD predicts that the dominant 2D spectrum is a $-5/3$ power law in $k_{\perp}$ i.e., $G^{\infty} (k_{\perp}) \equiv E^{\infty} (k_{\perp}) k_{\perp} = C^{\infty} k_{\perp}^{-5/3}$ \citep{Zank_etal_2017a, Zank_etal_2020a}, where $E^{\infty} (k_{\perp})$ is the 1D Els\"asser energy spectrum. However, because the flow is so highly magnetically aligned ($\Psi \simeq 3^{\circ}$ in the sub- and super-Alfv\'enic regions of interest), $P_{\parallel}^{\infty} (f)$ and $P_{\perp}^{\infty} (f)$ are effectively zero. Thus, as described above, the quasi-perpendicular fluctuations are essentially invisible to \textit{PSP} measurements unfortunately. 

\section{Observed Spectra and Theory}

\begin{figure}[htbp]
\begin{center}
$$\includegraphics[width= 0.5\textwidth]{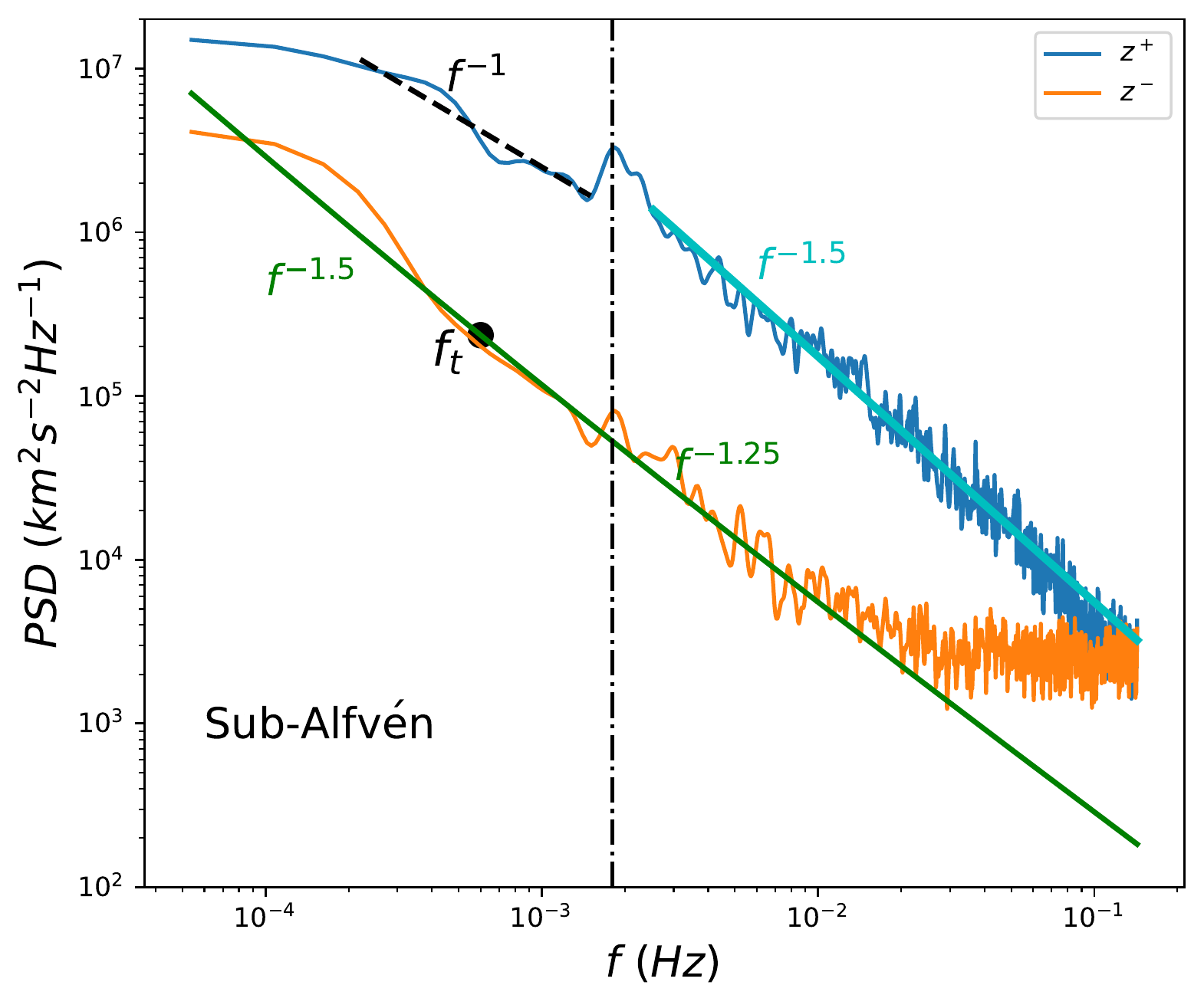}
\includegraphics[width= 0.5\textwidth]{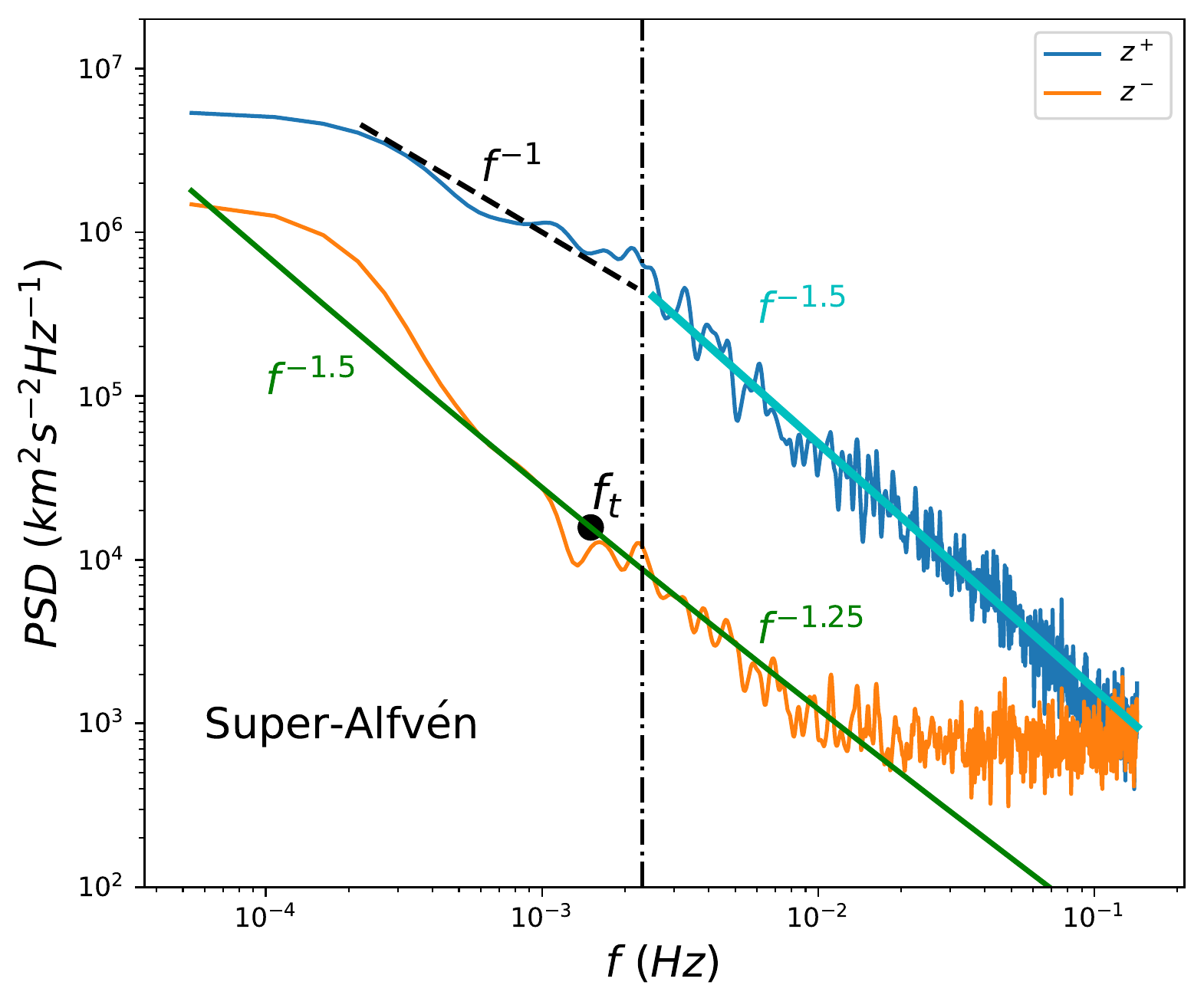}$$
\end{center}
\caption{  Trace spectra of the Els\"asser variables $\mathbf{z^\pm}$ calculated in 5-hour intervals for the sub-Alfv\'enic region (left panel), and a neighboring super-Alfv\'enic region (right panel). The sub-Alfv\'enic region is delineated by the two vertical dashed lines shown in Figure 1. The 5-hour interval just after the sub-Alfv\'enic region is selected as representative of the super-Alfv\'enic region. The solid green and cyan lines are predicted theoretical spectra, the dashed $f^{-1}$ curve is to guide the eye, and $f_t$ identifies the transition frequency (see text for details). The dashed vertical line separates the $f^{-1}$ and the $f^{-1.5}$ sections of the $\mathbf{z^+}$ spectra. } 
\label{fig:2}
\end{figure}

In Figure \ref{fig:2}, we plot the power spectral densities (PSDs) of the forward and backward Els\"asser variables ${\bf z}^{\pm}$ for the sub-Alfv\'enic region (9:33--14:42 UT) and a neighboring super-Alfv\'enic region (15:00--20:10 UT). This super-Alfv\'enic region was used because the discrepancy between the ion partial moment density measured by SPAN and the electron density estimated from FIELDS is large in the preceding super-Alfv\'enic region (see Figure \ref{fig:1}), due to part of the particle velocity distribution measured by SWEAP being blocked.
The PSD is evaluated over a 5-hour interval for each region. We use the standard Fourier method to calculate the trace spectra of $\mathbf{z^\pm}$ based on the Fourier-transformed autocorrelation function of the three components. The vertical dashed-dotted line in each panel denotes the frequency corresponding to the correlation scale that separates the energy-containing range and inertial range. A $f^{-1}$ spectrum for the energy-containing range is displayed as a reference. The cyan and green curves in each panel are the theoretical predicted spectra for $\mathbf{z^\pm}$ in both sub- and super-Alfv\'en regions. The dominant component in both regions is evidently the outward propagating $\mathbf{z^+}$ component, with the PSD being at least an order of magnitude larger than that of the inward $\mathbf{z^-}$ component. The flattening of the $\mathbf{z^-}$ PSD at frequencies $\ge 10^{-2}$ Hz has no physical significance and is due to the noise floor in the plasma measurements. Careful examination of the ${\bf z}^-$ PSD shows that the low frequency part of the spectrum is steeper than the high frequency part, whereas the ${\bf z}^+$ PSD, other than a bump at the inner scale \citep{Kasper_etal_2021}, is a single power law in frequency with $\sim f^{-1.5}$. The modestly super-Alfv\'enic spectra for ${\bf z}^{\pm}$ are very similar, and the dominance of ${\bf z}^+$ is again evident. 

Although not the focus of this work, we note that the dominant ${\bf z}^+$ spectrum exhibits a low frequency $f^{-1.0}$ power law, unlike the ${\bf z}^-$ spectrum. \cite{Matteini_etal_2018} offer an interesting explanation for why the dominant slab spectrum should exhibit an $f^{-1.0}$ spectrum and the minority not 
in terms of a saturation of the fluctuation amplitude at large scales imposed by the constraint ${\bf B} = \mbox{const.}$

 The values of $\Psi \equiv \langle \theta_{B_0 U_0} \rangle$ in the sub- and super-Alfv\'enic regions analyses in Figure \ref{fig:2} are $15^{\circ}$ and $18^{\circ}$ respectively. Unfortunately, both the strong alignment of mean magnetic field and mean velocity and the very limited range of $\theta_{B_0 U_0}$ in both intervals of interest, as shown in Figure \ref{fig:1}, make it very difficult to evaluate the ratio of 2D and slab power. To do so requires a wide range of angles $\theta_{B_0 U_0}$ ($\sim 0^{\circ} -  90^{\circ}$) and indeed \cite{Zhao_etal_2020a, Zhao_etal_2021, Wu_etal_2021} use the criterion $0^{\circ} < \theta_{B_0 U_0} < 20^{\circ}$ or $> 160^{\circ}$ to identify parallel intervals. We conclude that because of the parallel sampling in the two sub- and super-Alfv\'enic  intervals, the 2D component is not observed and hence does not contribute significantly to the observed slab component, not allowing us to assess accurately the 2D contribution. Further analysis is needed to clarify this, with hopefully a wider range of sampling angles observed in future sub-Alfv\'enic flows. 
From both the values of $\Psi$ and the normalized cross-helicity and residual energy spectrograms, it is clear that the spectra correspond to predominantly forward and minority backward propagating Alfv\'enic fluctuations since the 2D component is observationally invisible to \textit{PSP}. While one may argue that the $f^{-3/2}$ ${\bf z}^+$ spectrum is consistent with the Iroshnikov-Kraichnan theory, the argument is not credible since the needed counter-propagating ${\bf z}^-$ waves are almost entirely absent. The spectral theory of NI MHD \citep{Zank_etal_2020a} predicts that the general form of the NI/slab turbulence spectrum is given by 
\begin{equation}
G^* (k_z) \equiv E^*(k_z, k_{\perp}) k_{\perp}^2 = C^* k_z^{-(2a + 3)/3} \left( 1 + \left( \frac{k_z}{k_t} \right)^{-(2a - 3)/3} \right)^{1/2} , \label{eq:10}
\end{equation}
where $a$ describes a possible relationship between slab wave numbers $k_{\perp}$ and $k_z$, i.e., wave number anisotropy such that $k_{\perp} = k_z^a /k_t^{a-1}$ for $a > 0$. In deriving (\ref{eq:10}), as with any spectral theory, a crucial step is to identify the triple correlation time $\tau_3$ and then invoke a Kolmogorov phenomenology \citep{matthaeus_zhou_1989_extendedinertialrange, zhou_matthaeus_1990_mhdmodels, Zhou_etal_2004_review} for the NI/slab model, i.e., the NI/slab dissipation rate $\varepsilon_*$ satisfies $\varepsilon_* = \langle {z^*}^2 \rangle \tau_s = \tau_3 \langle {z^*}^2 \rangle /\tau_*^2$, where $\tau_s$ is the spectral transfer time given by $\tau_3/\tau_*^2$ and $\tau_*$ is the dynamical timescale $\tau_*^{-1} \equiv \langle {z^*}^2 \rangle^{1/2} k_{\perp} = \tau_3 \langle {z^*}^2 \rangle^2 k_{\perp}^2$.   \cite{Zank_etal_2020a} approximate the triple correlation time as the sum $\tau_3^{-1} = \tau_{\infty}^{-1} + \tau_A^{-1}$, where $\tau_{\infty}$ is the usual nonlinear timescale for the dominant 2D component, $\tau_{\infty}^{-1} = \langle {z^{\infty} }^2 \rangle^{1/2} /\lambda_{\perp}^{\infty}$, and $\lambda_{\perp}^{\infty}$ is the corresponding perpendicular correlation length.  A somewhat more complicated Alfv\'enic timescale $\tau_A$ is necessary \citep{Zank_etal_2020a} because the usual $V_A/\lambda_A$, $\lambda_A$ the Alfv\'enic correlation length, fails to capture two critical properties: 1) the Alfv\'en advection term does not contribute to nonlinear interactions or spectral transfer for unidirectionally propagating Alfv\'en waves, and 2) spectral transfer mediated by the Alfv\'en term is possible only when $\langle {z^{*+}}^2 \rangle \neq \langle {z^{*-}}^2 \rangle \neq 0$, i.e., for $|\sigma_c^*| \neq 1$, where $\sigma_c^*$ is the normalized NI/slab cross helicity. A suitable generalization of the Alfv\'en time scale that captures the two properties above is given by $\tau_A^{-1} = (V_{A0} /\lambda_A) {M_{A0}^t}^2 (1 - {\sigma_c^*}^2 )^{1/2}$ \citep{Zank_etal_2020a}\footnote{The parameter $M_{A0}^t = \langle u^2 \rangle^{1/2} / V_{A0}$ is the turbulent Alfv\'enic Mach number since  $\langle u^2 \rangle$ is the variance of the turbulent velocity fluctuations and represents the scaling parameter used in the NI MHD expansion in the $\beta_p \ll 1$ or $O(1)$ limits \citep{zank_matthaeus_1993_incompress, Zank_etal_2017a, Zank_etal_2020a}. The   presence of the term in the Alfv\'en timescale is a formal consequence of the NI scaling and has no additional physical significance.} where the subscript $0$ refers to mean magnetic field values. This then yields equation (\ref{eq:10}), which has a ``transition wave number'' $k_t$ or frequency $f_t$ (Figure \ref{fig:2}). The physical interpretation of $k_t$ is that it represents the transition from a wave number regime controlled primarily by nonlinear interactions to a regime controlled by Alfv\'enic interactions; specifically we have formally from the above definitions that $\tau_{\infty} /\tau_A = \left( k_z/k_t \right)^{-(2a - 3)/3}$ \citep{Zank_etal_2020a}. 

We can apply (\ref{eq:10}) to the spectra shown in Figure \ref{fig:2} by choosing $a = 3/4$, i.e., $k_{\perp} \sim k_z^{3/4}$. For the ${\bf z}^+$ spectrum, we assume that $\tau_{\infty} \ll \tau_A^+$, i.e., nonlinear interactions mediated by quasi-2D fluctuations dominate, and so obtain
\begin{equation}
G^{*+} (k_z) \simeq C^{*+} k_z^{-3/2}. \label{eq:11}
\end{equation}
This is reasonable given the absence of sufficient ${\bf z}^-$ modes with which to interact nonlinearly \citep{Dobrowolny_etal_1980a, Dobrowolny_etal_1980b}.
For the ${\bf z}^-$ spectrum, we suppose that $\tau_{\infty}$ and $ \tau_A$ are finite since the minor ${\bf z}^-$ component can interact with the more numerous counter-propagating ${\bf z}^+$ modes. Thus, there exists a transition wave number $k_t$ after which the ${\bf z}^-$ spectrum will flatten. Expression (\ref{eq:10}) becomes 
\begin{equation}
G^{*-} (k_z) = C^{*-} k_z^{-3/2} \left( 1 + \left( \frac{k_z}{k_t} \right)^{1/2} \right)^{1/2}, \label{eq:12}
\end{equation}
showing that at small wave numbers the spectrum is $\sim k_z^{-3/2}$ and at large wave numbers the asymptotic spectrum is $\sim k_z^{-1.25}$. We use equations (\ref{eq:6}) -- (\ref{eq:9}) to express the wave number spectra for the 2D component ($C^{\infty} k_{\perp}^{-5/3}$) and the slab component (\ref{eq:11}) and (\ref{eq:12}) in frequency space. The results are illustrated in Figure \ref{fig:2}. The left panel for the sub-Alfv\'enic region shows that the observed ${\bf z}^{\pm}$ PSDs are very well fitted by the theory for values of $C^{*\pm} = 0.13$ and $1.5$  respectively, and $k_t = 4.3 \times 10^{-4}$ km${}^{-1}$. The predicted flattening of the theoretical frequency spectrum $f^{-3/2}$ at higher frequencies to $f^{-1.25}$ fits the observed ${\bf z}^{-}$ PSD well. The super-Alfv\'enic interval spectra (right panel) are similarly well fitted with similar parameters {\bf $C^{*\pm} = 0.08$ and $0.012$ }respectively, and $k_t = 1.0 \times 10^{-4}$ km${}^{-1}$.  Since $\Psi$ is essentially parallel  in both intervals, the contribution from the 2D spectrum is very small and prevents us from evaluating $C^{\infty}$. The full anisotropy cannot therefore be determined because we cannot evaluate the power spectrum $P_{\perp}^{\infty} (f)$ (equations (\ref{eq:7}) and (\ref{eq:9})) observationally. The parameter $a$ relating $k_{\perp}$ and $k_z$ introduces a modest slab anisotropy such that $P_{\parallel}^* (k_z) > P_{\parallel}^* (k_{\perp})$, at sufficiently small scales. In the context of nearly incompressible MHD, this  implies the ordering $P_{\perp}^{\infty} (k_{\perp}) \gg P_{\parallel}^* (k_z) > P_{\parallel}^* (k_{\perp})$ with $a = 3/4$ \citep{Zank_etal_2020a}. As we discuss further below in the physical interpretation,  where observations of non-slab turbulence can be made by \textit{PSP}, a significant quasi-2D component has been observed \citep{Bandyopadhyay_McComas_2021, Zhao_etal_2021b}. 

The fitting parameters apply primarily to the spectral slopes and although the spectral indices for the $\mathbf{z^{\pm}}$ spectra are similar for both the super- and sub-Alfv\'enic intervals, there are some obvious differences. For example, the transition frequency $f_t$ shifts to a larger frequency in the super-Alfv\'enic region, suggesting that nonlinear interactions rather than Alfv\'enic interactions dominate more of the low-frequency spectrum. In addition, the spectral amplitude in the inertial range for both $\mathbf{z^{\pm}}$ in the sub-Alfv\'enic region is approximately 5 times larger than that in the super-Alfv\'enic region, and this appears to be true of the $f^{-1}$ energy-containing range for the $\mathbf{z^+}$ spectra too. 

\begin{figure}[htbp]
\begin{center}
\includegraphics[width= 0.5\textwidth]{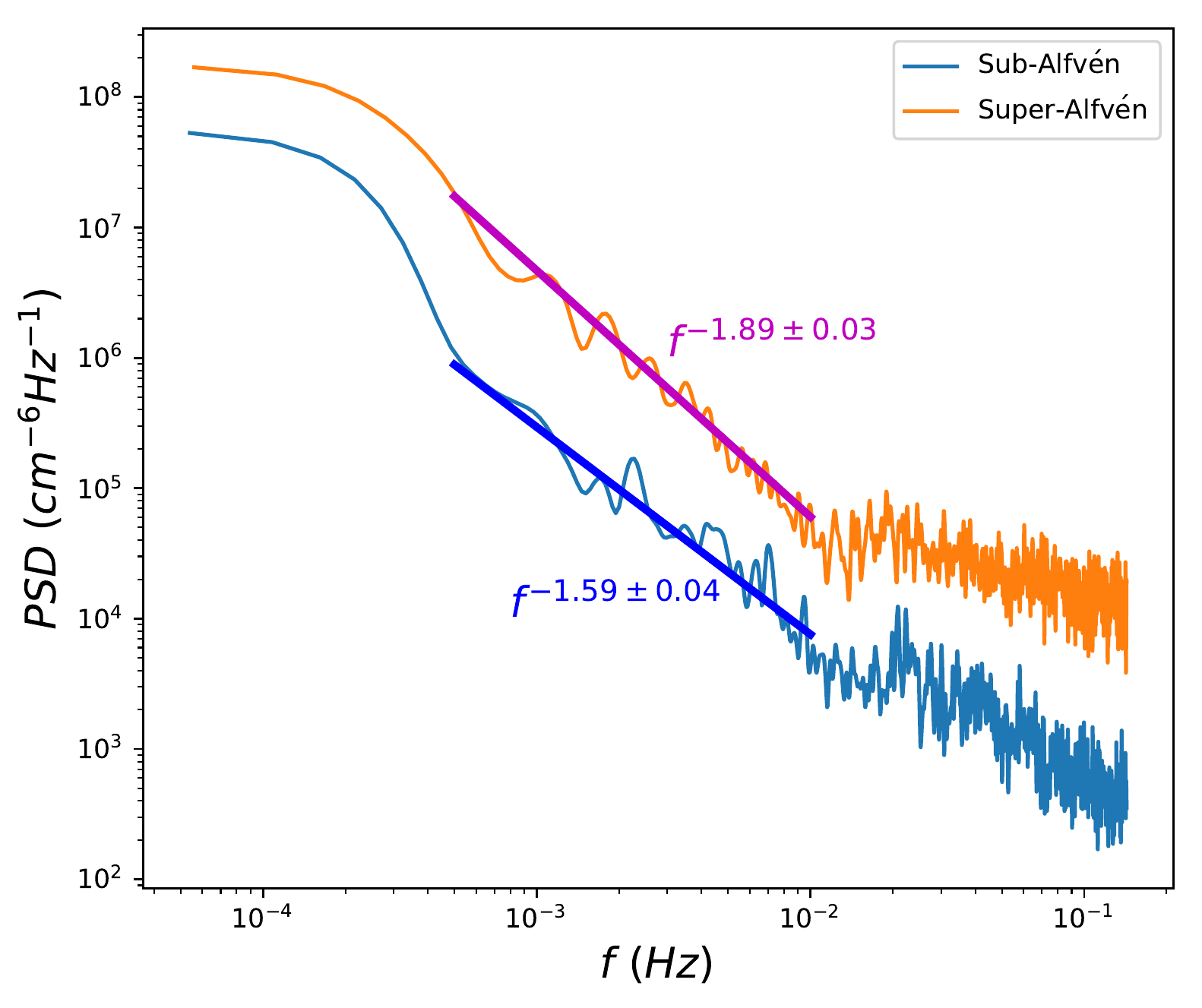}
\end{center}
\caption{ PSDs for the fluctuating density variance in the sub- and super-Alfv\'enic regions corresponding to those used for Figure \ref{fig:2}. The LFR electron density from FIELDS has been used. The fitted frequency range for both intervals starts from $5 \times  10^{-4}$ Hz to $1 \times 10^{-2}$ Hz. } 
\label{fig:3}
\end{figure}

Shown in Figure \ref{fig:3} are plots of the electron number density PSDs in the sub- and super-Alfv\'enic regions. Because the partial moment fractions for the PSP/SPAN-Ai data are entangled with the velocity fluctuations, it is difficult to evaluate the extent to which the density fluctuations are or are not passive structures embedded in incompressible turbulence.  For this reason, we use the more accurate low-frequency receiver (LFR) electron data set from the Radio Frequency Spectrometer (RFS) part of the PSP/FIELDS instrument suite \citep{Bale_etal_2016} since it is sufficiently well populated over the range shown in Figure \ref{fig:3}. The flattening at higher frequencies beyond some $10^{-2}$ Hz is likely real despite the LFR data being unreliable at these higher frequencies (an independent analysis calibrating the LFR measurements to the spacecraft floating voltage and using that as a proxy indicates a high frequency flattening followed by an eventual steepening of the electron density spectrum. This is not shown here since it is outside the low frequency inertial range of turbulence on which we are focussed.). 
The spectra over the interval $4 \times  10^{-4}$ Hz to $10^{-2}$ Hz are simple power laws with a spectral index of about $-1.59$ in the sub-Alfv\'enic flow and about $-1.89$ in the super-Alf\'enic flow. The density spectra for both regions do not correspond to the $\mathbf{z^+}$ spectra in the inertial range that have a $-1.5$ spectral index, and are quite unlike the convex $\mathbf{z^-}$ spectra (Figure \ref{fig:2}), thus ruling out the possibility that the density spectra are due to the parametric decay instability \citep{Goldstein_1978, Telloni_etal_2009, Bruno_etal_2014}.  
Density spectra can be determined from the NI MHD theory. Within the NI MHD theory, density fluctuations are entropic modes, i.e., zero frequency fluctuations, that are advected by the dominant quasi-2D turbulent velocity fluctuations and therefore behave as a passive scalar \citep{Zank_etal_2017a}. 
The relevant time scale is the quasi-2D nonlinear time scale and the underlying spectrum responsible for advection is that associated with the 2D velocity fluctuations. The NI MHD theory explicitly shows that the 2D Els\"asser variable spectrum satisfies $E_{2D} \sim k_{\perp}^{-5/3}$. Under some circumstances \citep{Zank_etal_2017a}, this serves as a proxy for the spectrum of 2D velocity fluctuations but more generally, the dominant quasi-2D velocity spectrum $E_{2D, v}$ will have the form  $E_{2D,v} \sim k_{\perp}^{-q}$. 
This yields a density spectrum of the form $E_{\rho} (k_{\perp}) = C_{\rho} k_{\perp}^{-q}$, where $q = 5/3$ only if e.g., kinetic energy dominates or the quasi-2D residual energy is zero. By extending the analysis of Section \ref{sec:frequency}, one can show that the frequency spectrum for the density PSD $P_{\rho} (f) $ is related to the density wave number PSD $P_{\rho} (k_{\perp})$ according to 
\begin{equation}
P_{\rho} (f) = C_{\rho} \int_{|k_x|}^{\infty} \frac{k_{\perp}^{-q} }{\sqrt{k_{\perp}^2 - k_x^2 } }dk_{\perp} = \frac{\sqrt{\pi} C_{\rho} k_x^{-q} \Gamma \left( \frac{q}{2} \right) }{2 \Gamma \left( \frac{q + 1}{2} \right) }, \quad k_x = \frac{2 \pi f}{U_0 \sin \Psi} , \label{eq:13}
\end{equation}
where $C_{\rho}$ is the amplitude of the density spectrum. The frequency spectrum therefore has the form $f^{-q}$. That the density spectrum is noticeably distinct from both the observed $\mathbf{z^{\pm}}$ spectra suggests its origin is unrelated to the slab spectra, whether via the parametric decay instability or passive advection of density fluctuations by slab turbulence. That leaves the possibilities that the density fluctuations are either zero frequency NI MHD entropic modes advected by quasi-2D incompressible turbulence \citep{Zank_etal_2017a} or compressible wave modes, which have been identified in \textit{PSP} data in the presence of dominant incompressible turbulence \citep{Zhao_etal_2021a}.  

\begin{figure}[htbp]
\begin{center}
$$\includegraphics[width= 0.5\textwidth]{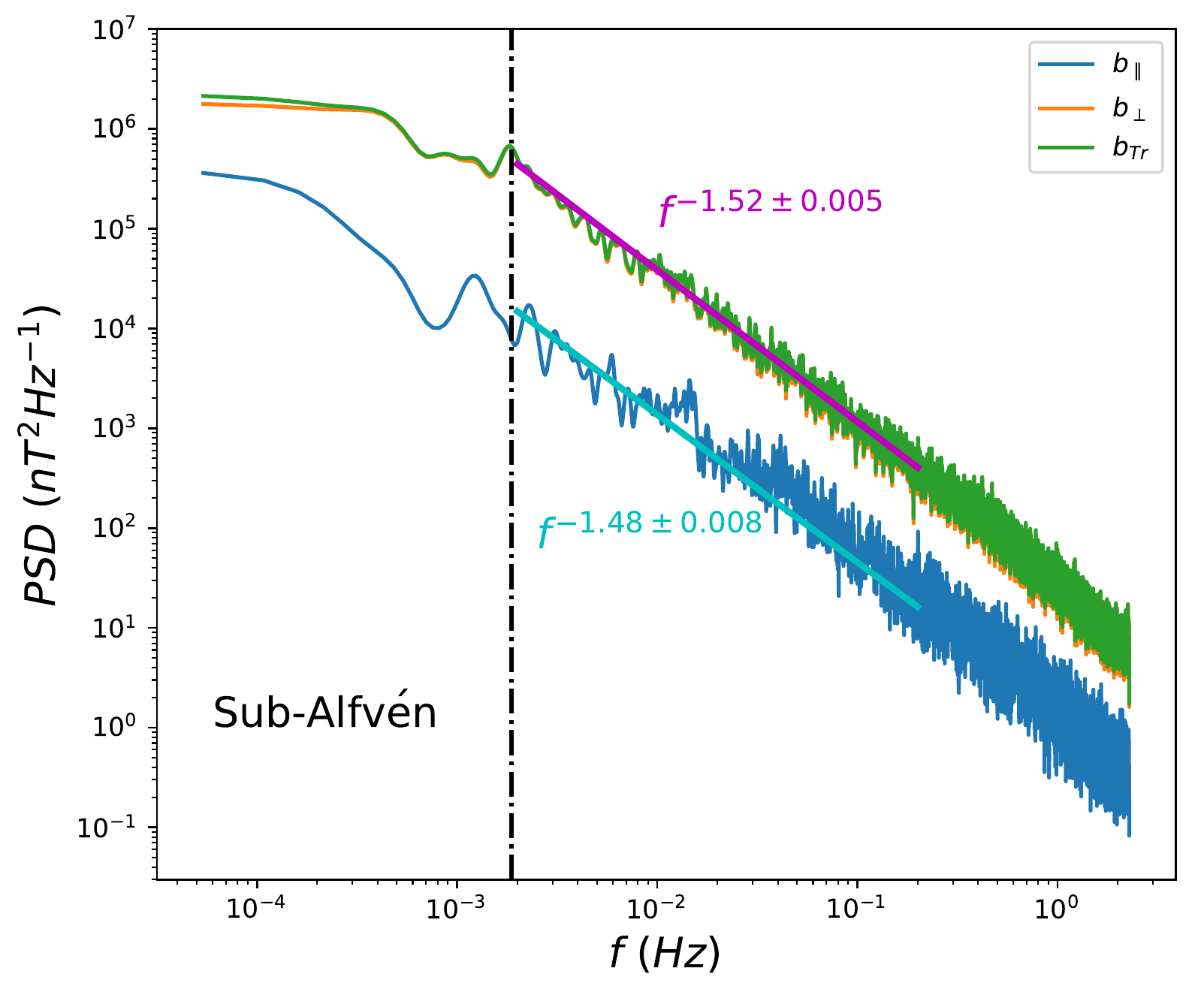}
\includegraphics[width= 0.5\textwidth]{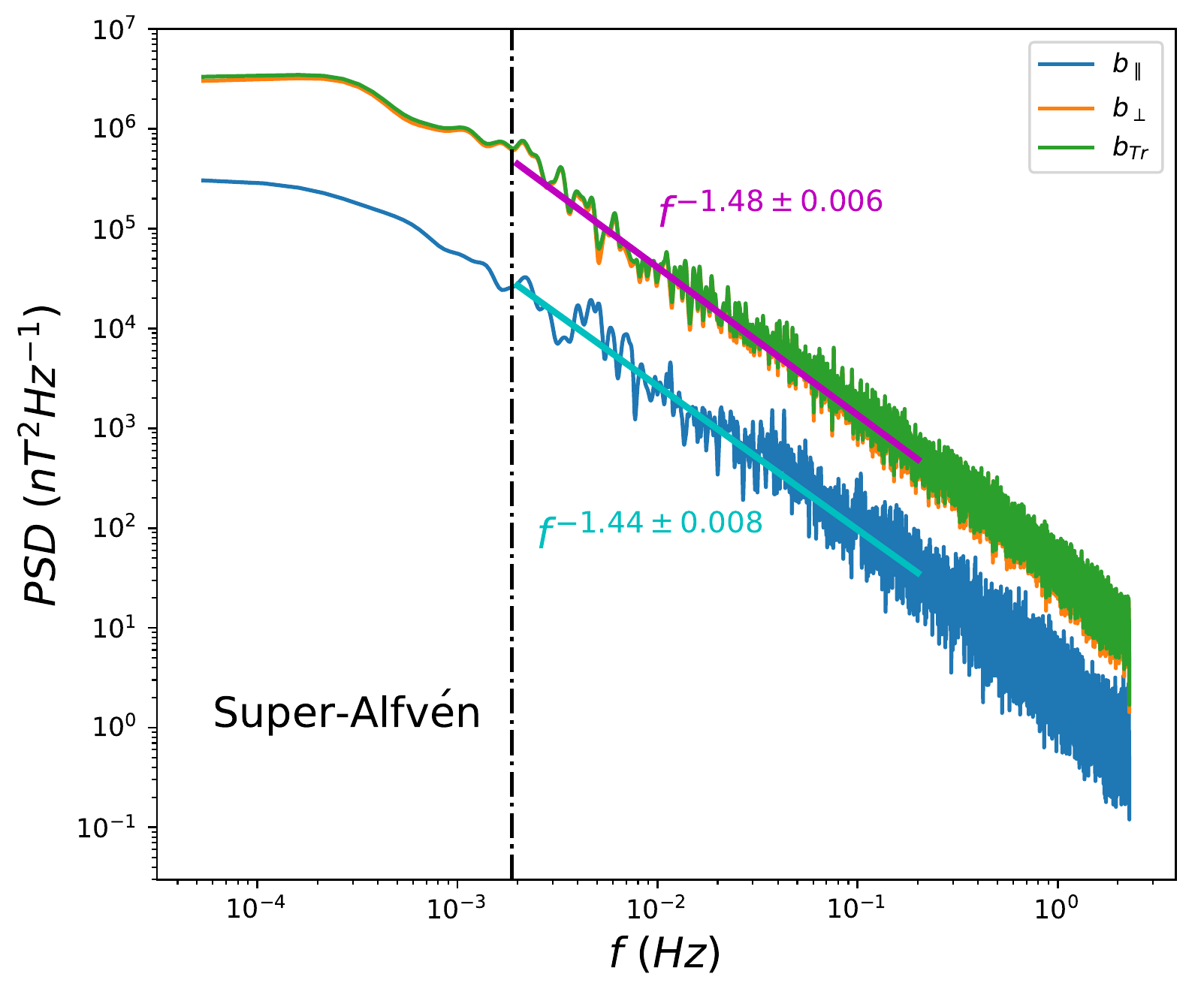}$$
\end{center}
\caption{ PSDs for magnetic fluctuations in the parallel (blue) and transverse (orange) directions for the sub-Alfv\'enic (left panel) and super-Alfv\'enic (right) regions. The total trace spectrum (green) effectively  overlays the transverse fluctuations. Pink and blue colored lines are fitted to the data  over the frequency range $2 \times 10^{-3}$ to $0.2$ Hz to estimate the spectral indices of the incompressible and compressible magnetic field PSDs. } 
\label{fig:4}
\end{figure}
The compressibility of the fluctuations is presented in Figure \ref{fig:4}, illustrating that the transverse or incompressible magnetic fluctuations, shown by the orange curve, are clearly dominant compared to the compressible magnetic field aligned fluctuations (blue curve). This is true of both sub-and super-Alfv\'enic regions and is consistent with observations discussed previously by \cite{Zhao_etal_2021a}. This assures us that the turbulence observed by PSP is largely incompressible, including in the sub-Alfv\'enic solar wind. The spectral slopes are found to be about $-1.52$ for incompressible magnetic fluctuations and about $-1.48$ for the corresponding compressible component in the sub-Alfv\'enic wind and respectively $-1.48$ and $-1.44$ in the super-Alfv\'enic interval over the frequency range $2 \times 10^{-3}$ to $0.2$ Hz. It is evident that the spectral indices of the observed density PSDs are quite different from those of the compressible spectra shown in Figure \ref{fig:4}. Whereas the compressible fluctuations shown in Figure \ref{fig:4} can be associated with fast (and possibly slow) magnetosonic modes \citep{Zhao_etal_2021a}, the very different characteristics of the density PSDs suggest that these fluctuations are not associated with waves but instead are likely to be entropy fluctuations with zero frequency. This result demonstrates \textit{post facto} the rationale for using Taylor's hypothesis for density fluctuations in the form of equation (\ref{eq:13}), i.e., for advected density fluctuations. Within NI MHD, density fluctuations are advected by the dominant quasi-2D turbulence in the $O(1)$ and $\ll 1$ plasma beta regimes \citep{hunana_zank_2010_apj, Zank_etal_2017a}, thereby providing insight into the dominant quasi-2D velocity turbulence that \textit{PSP} is unable to observe in the highly field-aligned intervals. In particular, the density spectrum in the sub-Alfv\'enic region suggest a quasi-2D velocity spectrum very slightly flatter than $-5/3$ and rather steeper than $-5/3$ in the super-Alfv\'enic interval.

Evidence for the presence of quasi-2D fluctuations has been presented elsewhere. \cite{Zhao_etal_2020, Zhao_etal_2020a, Zhao_etal_2021a} have identified small-scale flux ropes observed in previous \textit{PSP} encounters. \cite{Bandyopadhyay_McComas_2021} and \cite{Zhao_etal_2021b} find that turbulence in the inner heliosphere is highly anisotropic with significant contributions from a quasi-2D component, in many cases dominating the slab contribution despite the challenges for \textit{PSP} observing quasi-2D fluctuations when the magnetic field and plasma flow become increasingly highly aligned with decreasing distance above the solar surface. Accordingly, we can interpret the spectra illustrated in Figure \ref{fig:2} as due to the possibly minority slab component of quasi-2D+slab turbulence. This then yields the following physical interpretation of the observed ${\bf z}^{\pm}$ turbulent fluctuations in the sub-Alfv\'enic and modestly super-Alfv\'enic regions of the solar wind. The ${\bf z}^+$ (outward) fluctuations dominate, with the spectral amplitude of inward propagating slab modes nearly an order of magnitude smaller, meaning the slab component is comprised almost entirely of uni-directionally propagating Alfv\'en waves. Given the much smaller intensity of inward propagating modes, the ${\bf z}^+$ are obliged to interact nonlinearly almost exclusively with quasi-2D modes to produce the observed power law spectrum. The nonlinear interaction is governed by the nonlinear time scale $\tau_{\infty}$ with virtually no interaction on the Alfv\'en time scale $\tau_A$ with the counter-propagating minority ${\bf z}^-$ component. The spectral slope of $-3/2$ for the ${\bf z}^+$ (and low frequency part of the ${\bf z}^-$ spectrum) indicates modest {\bf slab} wave number anisotropy with $k_z \sim k_{\perp}^{4/3}$ in the inertial range. By contrast, the low-frequency ${\bf z}^-$ component is dominated by nonlinear rather than Alfv\'enic interactions, unlike the high frequencies that are governed primarily by interactions with counter-propagating Alfv\'en modes on the time scale $\tau_A^-$. This is manifest in the concavity of the ${\bf z}^-$ spectrum due to the presence of a transition wave number or frequency at which $\tau_{\infty} = \tau_A$. Nonetheless, to explain the slab observations presented here in the context of nearly incompressible MHD, the 2D component power anisotropy should dominate the power in the slab component. 

\section{Conclusions} \label{sec:conclusion}

\begin{enumerate}
\item The spectrograms for the normalized magnetic helicity $\sigma_m$, cross helicity $\sigma_c$, and residual energy $\sigma_r$ show that \textit{PSP} observed primarily outwardly propagating Alfv\'enic fluctuations during the first of the sub-Alfv\'enic intervals observed. This likely reflects the highly magnetic field-aligned flow of the interval that renders quasi-2D fluctuations effectively invisible to observations. Nonetheless, some evidence of magnetic structures is present near the interval boundaries, as well as a large vortex-like structure embedded in the interval.

\item We extended Taylor's hypothesis, allowing us to relate frequency and wave number spectra in our analysis of turbulence in sub-Alfv\'enic and the modestly super-Alfv\'enic flows, based on a decomposition of the turbulence into 2D and forward and backward propagating slab components. 

\item The PSDs for the ${\bf z}^{\pm}$ (Els\"asser) fluctuations were plotted for a sub- and super-Alfv\'enic interval, showing that the forward ${\bf z}^+$ component dominates, having a spectral amplitude much greater than that of the ${\bf z}^-$ PSD, and a frequency (wave number) spectrum of the form $f^{-3/2}$ ($k_{\parallel}^{-3/2}$) throughout the inertial range. By contrast, the ${\bf z}^-$ PSD exhibits a convex spectrum: $f^{-3/2}$ ($k_{\parallel}^{-3/2}$) at low frequencies that flattens around a transition frequency (wave number) $f_t$ ($k_t$) to $f^{-1.25}$ ($k_{\parallel}^{-1.25}$) at higher frequencies. Because \textit{PSP} makes measurements in a highly aligned flow, the observations correspond largely to slab fluctuations.

\item To interpret the observations, we apply a NI MHD 2D+slab spectral theory \citep{Zank_etal_2020a} to the ${\bf z}^{\pm}$ spectra, finding that the theoretically predicted slab  spectra are in excellent agreement with the observed spectra if there exists a modest {\bf slab} wave number anisotropy $k_{\perp} \sim k_{\parallel}^{3/4}$. The ${\bf z}^+$ wave number spectrum is predicted to be $k_{\parallel}^{-3/2}$ because it interacts primarily with quasi-2D fluctuations on a time scale $\tau_{\infty}$ rather than the significantly smaller ${\bf z}^-$ component. By contrast, the minority ${\bf z}^-$ fluctuations can interact with both quasi-2D and counter-propagating slab modes, so that both the nonlinear $\tau_{\infty}$ and Alfv\'en $\tau_A$ time scales determine the form of the spectrum. Theoretically, this combination of time scales predicts a convex spectrum with the inflection or transition point determined by the balance of the time scales, $\tau_{\infty} = \tau_A$, and the spectrum is predicted to flatten from a $f^{-3/2}$ ($k_{\parallel}^{-3/2}$) low frequency or nonlinear dominated regime to a $f^{-1.25}$ ($k_{\parallel}^{-1.25}$) higher frequency or Alfv\'enic dominated regime. Plots of the transverse and compressible magnetic field fluctuations show that turbulence in the {sub- and modestly super-Alfv\'enic flows are}  dominated by incompressible fluctuations. 

\item  The PSDs for the density fluctuations were plotted for both intervals of interest, exhibiting  simple power laws with spectral indices of $-1.59$ and $-1.89$ for the sub-and super-Alfv\'enic cases, respectively. The spectra do not resemble either the dominant $\mathbf{z^+}$ PSD and are distinctly different from the convex structured $\mathbf{z^-}$ spectrum, suggesting that the density fluctuations are not due to the parametric decay instability. The compressible magnetic field fluctuation spectrum follows the incompressible magnetic field spectrum closely and is distinctly different from the density spectrum, suggesting that the density fluctuations are not primarily compressible magnetosonic wave modes. Instead, they appear to be zero-frequency entropic modes advected by the background turbulent velocity field. This interpretation is consistent with the expectations of NI MHD in which entropic density fluctuations are advected by the dominant quasi-2D velocity fluctuations, indicating that the density spectra offer insight into quasi-2D turbulence. 

\item We find that the spectra in the neighboring modestly super-Alfv\'enic regions closely resemble those in the sub-Alfv\'enic interval, and indeed the three parameters ($C^{*\pm}$, and $a$) for the two sets of spectra are very similar, indicating that the same basic turbulence physics holds in both regions. Nonetheless, there are some differences in details, such as the transition frequency shifting to a larger frequency in the super-Alfv\'enic region, and the fluctuating power for both $\mathbf{z^+}$ and $\mathbf{z^-}$ in the sub-Alfv\'enic region is approximately 5 times larger than that in the super-Alfv\'enic region. The larger $k_t$ for the super-Alfv\'enic flow indicates that nonlinear interactions rather than Alfv\'enic interactions dominate for a larger part of the low-frequency spectrum. 

\item The physical interpretation of the Els\"asser forward and backward slab and density spectra reflects a manifestation of dominant quasi-2D turbulent  fluctuations in the solar wind. The same parameters explain both the observed forward and backward Els\"asser spectra. Since the fitting of the results is predicated on a quasi-2D nonlinear time scale and a quasi-2D spectrum of the Kolmogorov form, the results presented here suggest the presence of a dominant 2D component that, because of the field-aligned sampling in both intervals during this encounter, cannot be observed by \textit{PSP}, but nevertheless controls the evolution of slab and density turbulence in the sub-Alfv\'enic solar wind. 

\end{enumerate}



\acknowledgments
GPZ, LLZ, and LA acknowledge the partial support of a NASA Parker Solar Probe contract SV4-84017, an NSF EPSCoR RII-Track-1 Cooperative Agreement OIA-1655280, a NASA IMAP subaward under NASA contract 80GSFC19C0027, and a NASA award 80NSSC20K1783. DT was partially supported by the Italian Space Agency (ASI) under contract 2018-30-HH.0. \textit{Parker
Solar Probe} was designed, built, and is now operated by the Johns Hopkins Applied Physics Laboratory as part of NASA’s Living with a Star (LWS) program (contract NNN06AA01C). Support from the LWS management and technical team has played a critical role in the success of the \textit{Parker Solar Probe} mission. 



\bibliography{references}{}
\bibliographystyle{aasjournal}



\end{document}